\documentclass[a4paper,twoside]{article}
\newcommand{\affil}[1]{$^{\rm #1}$}
\usepackage[authoryear]{natbib}
\bibpunct{(}{)}{;}{a}{}{,}
\usepackage{graphicx}
\date{} 

\title{\large\bf\flushleft Chemical Evolution of Heavy Elements in the Early
Galaxy: Implications for Stellar Sources}
\author{\parbox{\textwidth}{\flushleft
\vspace{-0.5cm}
{\it Yong-Zhong Qian\affil{A,B}}\\
\vspace{0.4cm}
{\small \affil{A}\,School of Physics and Astronomy, University of Minnesota, 
Minneapolis, MN 55455, USA}\\
{\small \affil{B}\,Email: qian@physics.umn.edu}}}

\begin{document}
\twocolumn[
\begin{changemargin}{.8cm}{.5cm}
\begin{minipage}{.9\textwidth}
\vspace{-1cm}
\maketitle
\small{\bf Abstract:}
An overview of the sources for heavy elements in the early Galaxy is given.
It is shown that observations of abundances 
in metal-poor stars can be used along with a basic understanding of stellar models 
to guide the search for the source of the heavy $r$-process nuclei ($r$-nuclei). 
Observations show that this source produces very little 
of the elements from C through Zn including Fe. This strongly suggests that 
O-Ne-Mg core-collapse supernovae (SNe) from
progenitors of $\sim 8$--$11\,M_\odot$ are the source for the heavy $r$-nuclei.
It is shown that a two-component model based on the 
abundances of Fe (from Fe core-collapse SNe) and 
Eu (from O-Ne-Mg core-collapse SNe) gives very good 
quantitative predictions for the abundances of all the other elements in 
metal-poor stars.
 

\medskip
\medskip
\end{minipage}
\end{changemargin}
]
\small

\section{Introduction}
There are two approaches to study chemical evolution. The straightforward
approach relies on the knowledge of all nucleosynthetic sources, their
elemental yields, their occurrences in the interstellar medium (ISM), and 
their exchange of material with the ISM. Based on this knowledge, both
the average and the scatter for the abundance of any element in the ISM
can be calculated as functions of time. Unfortunately, our knowledge of
nucleosynthetic sources is rather incomplete. So this forward approach
cannot be applied to all elements. On the other hand, the elemental
abundances in stars generally reflect the composition of the ISM from
which they formed. Given a large collection of data on stellar abundances,
it is possible to infer the characteristics of some nucleosynthetic sources
from these data with the help of some basic understanding of stellar 
models. This reverse approach works best at early times when only a small
number of potential sources could have contributed to the ISM. This
paper presents a combination of forward and reverse approaches
to study chemical evolution of heavy elements in the early Galaxy.
The goal is to illustrate what can be learned about stellar sources from 
such studies.

Here ``early'' means the first Gyr or so of Galactic history. During this
epoch, only short-lived massive stars could have had time to evolve and
contribute their nucleosynthetic products to the ISM. As $\sim 1/3$ of
the solar Fe inventory was contributed by core-collapse supernovae 
(SNe, mostly Type II)
associated with massive stars over a period of $\sim 10$~Gyr, ``early
times'' also correspond to $<1/30$ of the solar Fe abundance
in the ISM. 
In the notation $[{\rm Fe/H}]\equiv\log({\rm Fe/H})-\log({\rm Fe/H})_\odot$,
such times correspond to metal-poor ISM with $[{\rm Fe/H}]<-1.5$.
Three groups of heavy elements will be discussed: (1) the elements
from C through Zn with mass numbers $A<70$, (2) the elements
from Sr through Ag with $A\sim 88$--110, and (3) the elements with
$A\sim 130$ through the actinides attributed to the rapid neutron 
capture process (the $r$-process). The solar abundances of the 
elements with $A\sim 88$--209 received important contributions
from the slow neutron capture process (the $s$-process). However,
as the major $s$-processing occurs in long-lived (with lifetimes
of $>1$~Gyr) stars of a few
$M_\odot$, the $s$-process contributions to the ISM
at $[{\rm Fe/H}]<-1.5$ can be ignored to good approximation.
Likewise, contributions to the Fe group elements from Type Ia SNe 
associated with white dwarfs evolved from long-lived low-mass stars 
in binaries can also be ignored at $[{\rm Fe/H}]<-1.5$.

\section{Elements from C through Zn}
The relevant sources for the elements from C through Zn
are massive stars of $>11\,M_\odot$,
which give rise to Fe core-collapse SNe at the end of their evolution. 
The elements from C through Al are mainly produced by hydrostatic burning 
during the pre-SN evolution, and the elements from Si through Zn are
mainly produced by explosive burning associated with the propagation
of the SN shock through the shells above the Fe core. While the production
of all these elements has some important dependence on the initial 
metallicity of the SN progenitor through e.g., mass loss, pre-SN density
structure, and details of the explosion, theoretical yields of individual elements
are broadly similar over a wide range of metallicities and have been 
calculated by several groups (e.g., \citealt{ww95,chieffi,tominaga}).
There is an approximate overall agreement between the theoretical
yields averaged over the mass distribution of SN progenitors and
the data (e.g., \citealt{cayrel}) on the abundances of C through Zn in metal-poor 
stars with $-4<{\rm [Fe/H]}<-3$ (e.g., \citealt{frohlich1,tominaga}).
However, there are also important deficiencies of the SN models. 
For example, the calculated abundance ratios of N, K, Sc, Ti, Mn, 
and Co relative to Fe are too low compared with observations
(e.g., \citealt{tominaga}). Nitrogen is produced by the CN cycle 
when the C from He burning is mixed into the H burning shell.
This production can be greatly enhanced by rotationally-induced
mixing when rotation is explicitly included in evolutionary models 
of metal-poor massive stars (e.g., \citealt{meynet}). 
On the other hand, the underproduction of
K, Sc, Ti, Mn, and Co can be remedied by modifying e.g.,
the electron fraction $Y_e$ of the material undergoing
explosive nucleosynthesis.

The electron fraction $Y_e$ specifies the neutron-to-proton ratio 
of the material and plays a crucial role in nucleosynthesis.
The conversion between neutrons and protons can only proceed
through the weak interaction involving neutrinos. The death of
massive stars can be considered as a neutrino phenomenon.
When the Fe core of such a star collapses into a protoneutron star,
a great amount of gravitational binding energy is released in
$\nu_e$, $\bar\nu_e$, $\nu_\mu$, $\bar\nu_\mu$, $\nu_\tau$,
and $\bar\nu_\tau$ with average energies of
$\langle E_\nu\rangle\sim 10$--20~MeV. 
Typical luminosity of the initial neutrino
emission is $L_\nu\sim 10^{52}$~erg/s per species. In the case
of stable protoneutron stars,
neutrino emission with $L_\nu\sim 10^{51}$~erg/s 
per species lasts for $\sim 20$~s. With such intense neutrino
fluxes, the neutron-to-proton ratio of the material close to the
protoneutron star is modified by the reactions
\begin{eqnarray}
\bar\nu_e+p&\to& n+e^+,\label{eq-anuep}\\
\nu_e+n&\to& p+e^-.\label{eq-nuen}
\end{eqnarray} 
The $Y_e$ relevant for explosive nucleosynthesis in this material
then depends on the competition between the above two reactions
(e.g., \citealt{qian1,fuller,qwo}). The rates of these reactions are
proportional to the neutrino flux $L_\nu/\langle E_\nu\rangle$
and the cross section, which in turn is proportional to
$\langle(E_\nu\mp\Delta)^2\rangle\approx
\langle E_\nu^2\rangle\mp 2\Delta\langle E_\nu\rangle$
with the minus sign being for reaction~(\ref{eq-anuep}), the
plus sign for reaction~(\ref{eq-nuen}), and $\Delta=1.293$~MeV 
being the neutron-proton mass difference. When the reaction rates are
sufficiently high, $Y_e$ is determined by their ratio as 
\citep{qwo}
\begin{equation}
Y_e\approx\left[1+\frac{L_{\bar\nu_e}}{L_{\nu_e}}\left(
\frac{\epsilon_{\bar\nu_e}-2\Delta}{\epsilon_{\nu_e}+2\Delta}\right)
\right]^{-1},
\label{eq-ye}
\end{equation}
where $\epsilon_\nu\equiv\langle E_\nu^2\rangle/\langle E_\nu\rangle$.

\begin{figure}[h]
\begin{center}
\includegraphics[scale=0.35]{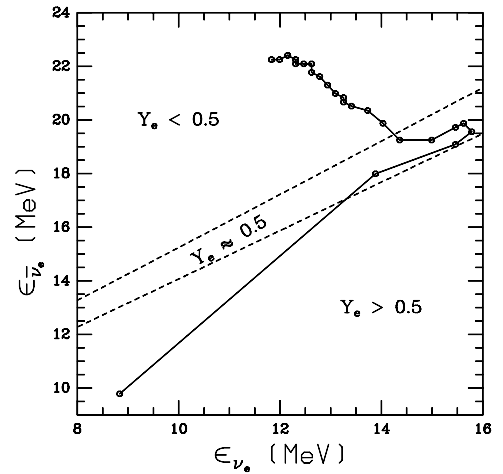}
\caption{Example evolution of $\epsilon_{\bar\nu_e}$ and 
$\epsilon_{\nu_e}$ for $\bar\nu_e$ and $\nu_e$ emitted by
a protoneutron star (solid curve). Time increases along the
solid curve starting from the lower left end. The ratio
$L_{\bar\nu_e}/L_{\nu_e}$ stays close to 1 throughout the
evolution. Values of
$\epsilon_{\bar\nu_e}$ and $\epsilon_{\nu_e}$ lying above
the upper dashed line would give $Y_e<0.5$ for
$L_{\bar\nu_e}/L_{\nu_e}=1$, those below the lower
dashed line would give $Y_e>0.5$ for
$L_{\bar\nu_e}/L_{\nu_e}=1.1$, and those between the two
dashed lines would give $Y_e\approx 0.5$ with the exact
$Y_e$ being sensitive to the value of $L_{\bar\nu_e}/L_{\nu_e}$.
See \citet{qwo} for details.}
\label{fig-enu}
\end{center}
\end{figure}

As an example, the pairs of $\epsilon_{\bar\nu_e}$ and $\epsilon_{\nu_e}$ 
for $\bar\nu_e$ and $\nu_e$ emitted by the protoneutron star 
in the SN model of \citet{woosley} during the first $\sim 20$~s of its life 
are shown as the solid curve in Fig.~\ref{fig-enu}. 
The lower left end of the curve corresponds to the onset of the Fe core 
collapse at time $t\approx 0$. Time increases along the curve in intervals
of $\approx 1/3$~s for $t\approx 0$--4~s and $\approx 1$~s for $t>4$~s.
The ratio $L_{\bar\nu_e}/L_{\nu_e}$ stays close to 1 throughout the
evolution. Values of $\epsilon_{\bar\nu_e}$ and $\epsilon_{\nu_e}$ lying 
above the upper dashed line in Fig.~\ref{fig-enu} would give $Y_e<0.5$ for 
$L_{\bar\nu_e}/L_{\nu_e}=1$ according to Eq.~(\ref{eq-ye}), those
below the lower dashed line would give $Y_e>0.5$ for 
$L_{\bar\nu_e}/L_{\nu_e}=1.1$, and those between the two
dashed lines would give $Y_e\approx 0.5$ with the exact
$Y_e$ being sensitive to the value of $L_{\bar\nu_e}/L_{\nu_e}$. 

During the first $\sim 1$~s after the onset of Fe core collapse, the
neutrino emission characteristics $L_{\bar\nu_e}\approx L_{\nu_e}$
and $\epsilon_{\bar\nu_e}<\epsilon_{\nu_e}+4\Delta$ (i.e., below
the upper dashed line in Fig.~\ref{fig-enu}) give rise to $Y_e>0.5$ in 
the material immediately above the protoneutron star. This results in
greatly-enhanced yields of $^{45}$Sc, $^{49}$Ti, and $^{64}$Zn
due to the production of their more proton-rich progenitor nuclei
$^{45}$Cr, $^{45}$V, $^{49}$Mn, and $^{64}$Ge \citep{pruet1,frohlich1}.
In particular, it was shown that the inclusion of neutrino effects on $Y_e$ 
dramatically increases the calculated abundance ratios Sc/Fe and Zn/Fe 
to the level in accordance with observations of metal-stars with 
$-4.1<[{\rm Fe/H}]<-0.8$ \citep{frohlich1}. Therefore, it appears
that Fe core-collapse SNe from progenitors of $>11\,M_\odot$
are reasonably established as the major source for the elements from
C through Zn in the early Galaxy.

\section{Production of Heavy Elements in the Neutrino-Driven Wind}
Reactions~(\ref{eq-anuep}) and (\ref{eq-nuen}) 
not only are important for determining the $Y_e$ of the material
close to the protoneutron star, but also heat this material,
enabling it to expand away from the protoneutron star as 
a neutrino-driven wind (e.g., \citealt{qwo}). In fact, as long as a 
stable protoneutron star is formed by some core collapse,
its neutrino emission drives such a wind for $\sim 20$~s.
As the wind expands away from the protoneutron 
star, neutrinos can continue to affect nucleosynthesis. For example,
the nuclear flow in proton-rich conditions encounters bottlenecks at 
nuclei with extremely slow proton-capture 
and $\beta^+$-decay rates. In the presence of an intense $\bar\nu_e$
flux, the neutrons produced by reaction~(\ref{eq-anuep})
can be captured by such nuclei to break through the bottleneck,
giving rise to the so-called $\nu p$-process \citep{frohlich2}, which
can produce many nuclei beyond $^{64}$Zn. It is important to 
find out whether the observed correlation between the abundances 
of Ge and Fe \citep{cowan2} can be accounted for after the 
contribution to Ge from the $\nu p$-process is taken into account.
General studies of nucleosynthesis in the proton-rich neutrino-driven 
wind were carried out in detail by 
\citet{frohlich2}, \citet{pruet2}, and \citet{wanajo2}.

As shown in Fig.~\ref{fig-enu}, the difference between
$\epsilon_{\bar\nu_e}$ and $\epsilon_{\nu_e}$ increases with time
and the neutrino-driven wind eventually becomes neutron-rich 
(i.e., $Y_e<0.5$). What can be produced in a neutron-rich wind?
In general, nucleosynthesis during expansion
of material from an initial state of high temperature depends on
$Y_e$, the entropy $S$, and the dynamic expansion
timescale $\tau_{\rm dyn}$ of this material. In the neutrino-driven
wind, these three parameters are determined by the mass, radius,
and neutrino emission characteristics of the protoneutron star
(e.g., \citealt{qwo}).  As the neutrino emission characteristics 
evolve, the conditions in the wind change accordingly. For illustration,
we consider two sets of conditions corresponding to winds taking off 
at two different times from a protoneutron star of $1.4\,M_\odot$
with  a radius of 10~km. At an earlier time, the wind has
$Y_e\approx 0.47$, $S\approx 70$ (in units of Boltzmann constant
per nucleon), and $\tau_{\rm dyn}\approx 0.024$~s. As the neutron-rich
material in this wind expands, the free nucleons first combine into
$\alpha$-particles. This consumes essentially all the protons. Then
an $\alpha$-process occurs to burn $\alpha$-particles and the remaining
neutrons into heavier nuclei \citep{hoffman1}. By the time all 
charged-particle reactions (CPRs) cease at sufficiently low temperature
due to the Coulomb barrier, the dominant products are Sr, Y, and Zr
with $A\sim 90$ and no neutrons are left to further process these nuclei
by neutron capture \citep{hoffman2}. For a late wind with
$Y_e\approx 0.37$, $S\approx 90$, and $\tau_{\rm dyn}\approx 0.066$~s,
the dominant products are Zr, Mo, Ru, Rh, Pd, and Ag
with $A\sim 96$--110 \citep{hoffman2}.
These nuclei are also produced by CPRs during
the $\alpha$-process and no neutrons are left at the end. As another
example, we consider a late wind from a heavier protoneutron star of 
$2\,M_\odot$ but with the same radius of 10~km. The conditions of
$Y_e\approx 0.35$, $S\approx 140$, and $\tau_{\rm dyn}\approx 0.11$~s
in this wind result in major production of Sn, Sb, and Te with $A\sim 124$ 
in addition to the nuclei with $A\sim 96$--110. Once again, all these nuclei 
are produced by CPRs during the $\alpha$-process and few neutrons are 
left at the end \citep{hoffman2}.

The above discussion shows that for the typical conditions in a
neutron-rich wind, elements from Sr through Ag with $A\sim 88$--110,
and in some cases nuclei with $A\sim 124$, are produced by CPRs
during the $\alpha$-process. As no or very few neutrons are left at
the end of the $\alpha$-process, these nuclei are the main products
in such winds. In contrast, if the abundance ratio of neutrons to heavy 
nuclei greatly exceeds $\sim 10$ when CPRs cease to occur at a 
temperature of several $10^9$~K due to the Coulomb barrier, 
then the $\alpha$-process smoothly merges
with the $r$-process as the heavy nuclei produced by the former
rapidly capture neutrons at lower temperatures. This is the
neutrino-driven wind model for the $r$-process (e.g.,
\citealt{woba,meyer1,takahashi,woosley,wanajo3}). 
In this model, the final abundance pattern produced by the $r$-process 
depends on how many neutrons are left for each seed nucleus at the 
end of the $\alpha$-process. For example, a neutron-to-seed ratio
of $n/s\sim 40$ would dominantly produce an abundance peak at 
$A\sim 130$ while another value of $n/s\sim 90$ would dominantly
produce nuclei of $A>130$ with an abundance peak at $A\sim 195$.

In general, a specific value of $n/s$ can be achieved by various
combinations of $Y_e$, $S$, and $\tau_{\rm dyn}$. A lower $Y_e$
corresponds to a higher initial neutron abundance and usually also
means that more neutrons are left at the end of the $\alpha$-process.
A value of $S>10$ indicates that the energy density of the material 
is dominated by radiation (and electron-positron pairs for sufficiently 
high temperature). The production
of seed nuclei is severely suppressed for $S>100$. This is because
in producing the seed nuclei the nuclear flow must rely on the 
three-body reaction $\alpha+\alpha+n\to{^9{\rm Be}}+\gamma$
to bridge the gap at $A=5$ and 8. It only requires a photon
of 1.573~MeV to dissociate a $^9$Be nucleus
back into two $\alpha$-particles
and a neutron. For a high $S$, there are a significant number of 
such photons in the high-energy tail of the Bose-Einstein distribution
over the temperature range of $\sim 3$--$6\times 10^9$~K for
the $\alpha$-process. The dynamic timescale $\tau_{\rm dyn}$
controls how fast the temperature drops, thereby specifying the
duration of the $\alpha$-process. Clearly, the shorter $\tau_{\rm dyn}$
is, the fewer seed nuclei are produced. In summary, a lower $Y_e$,
or a higher $S$, or a shorter $\tau_{\rm dyn}$ tends to give a larger 
$n/s$.

\begin{figure}[h]
\begin{center}
\includegraphics[scale=0.35, angle=270]{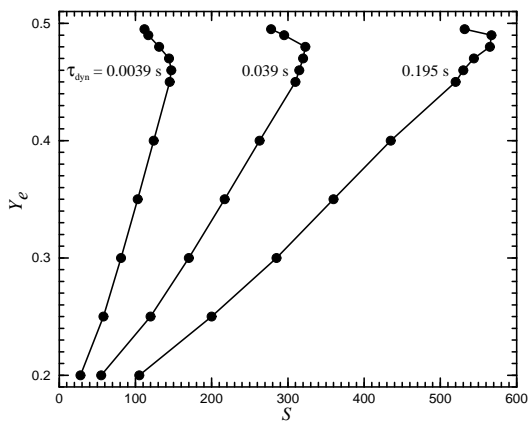}
\caption{Combinations of $Y_e$ and $S$ that would result in major 
production of an $r$-process abundance peak at $A\sim 195$ 
during expansion of material from an initial state of high temperature
for three values of $\tau_{\rm dyn}$. See \citet{hoffman2} for details.}
\label{fig-syetau}
\end{center}
\end{figure}

Combinations of $Y_e$ and $S$ that would result in major production
of an $r$-process abundance peak at $A\sim 195$ for three different
values of $\tau_{\rm dyn}$ (\citealt{hoffman2}; see also 
\citealt{meyer2,freiburghaus1})
are shown in Fig.~\ref{fig-syetau}. It can be seen that
for $Y_e\sim 0.4$ and $S\sim 100$ typical of the neutrino-driven
wind, an extremely short dynamic timescale of 
$\tau_{\rm dyn}\sim 0.004$~s is required
to produce an $r$-process abundance peak at $A\sim 195$.
However, the actual dynamic timescale typical of the 
wind is $\tau_{\rm dyn}\sim 0.01$--0.1~s. Studies by 
several groups \citep{qwo,witti,thompson1} found that
it is generally very difficult to obtain the conditions required
for producing an $r$-process abundance peak at $A\sim 195$
in the neutrino-driven wind, although it is quite plausible that 
the conditions in the wind from at least some protoneutron
stars (e.g., those with masses of $\sim 2\,M_\odot$)
are sufficient for producing $r$-process nuclei up to $A\sim 130$.

\section{Source for $r$-Process Nuclei with $A>130$}
In search of the source for the heavy $r$-process nuclei ($r$-nuclei)
with $A>130$, especially those in the peak at $A\sim 195$
of the solar $r$-process abundance pattern ($r$-pattern; e.g., 
\citealt{arlandini}), there were attempts to modify the conditions 
in the neutrino-driven wind by e.g., including the effects of a magnetic 
field above the protoneutron star \citep{thompson2} as well as proposals
of alternative sites such as neutron star mergers 
(e.g., \citealt{freiburghaus2}) and
the wind from the accretion disk around a black hole 
\citep{pruet3,mclaughlin}. Yet another approach is to seek 
guidance from observations. For example, the detection of
the $r$-process element ($r$-element) 
Ba with $A\sim 135$ in a number
of stars with [Fe/H]~$<-3$, especially the high Ba enrichments
in several stars with [Fe/H]~$\sim -3$, is crucial in evaluating  
neutron star mergers as the major source for the heavy 
$r$-nuclei. These events are much rarer 
(by at least a factor of $10^3$) than Fe core-collapse SNe. 
If neutron star mergers were the major source for the heavy 
$r$-nuclei, then enrichment in these nuclei would 
not occur until the ISM had already been substantially enriched in
Fe by Fe core-collapse SNe \citep{qian2,argast}. This is in contradiction 
to the observations of stars with significant to high Ba abundances
but very low Fe abundances. Therefore, it appears very unlikely
that neutron star mergers are the major source for the heavy 
$r$-nuclei. In contrast, if some events with a Galactic rate of 
occurrences similar to that of Fe core-collapse SNe are the source 
for such nuclei, then the expected relationship between Ba and Fe 
abundances is in good agreement with stellar observations 
(e.g., \citealt{argast}).

\begin{figure}[h]
\begin{center}
\includegraphics[scale=0.4]{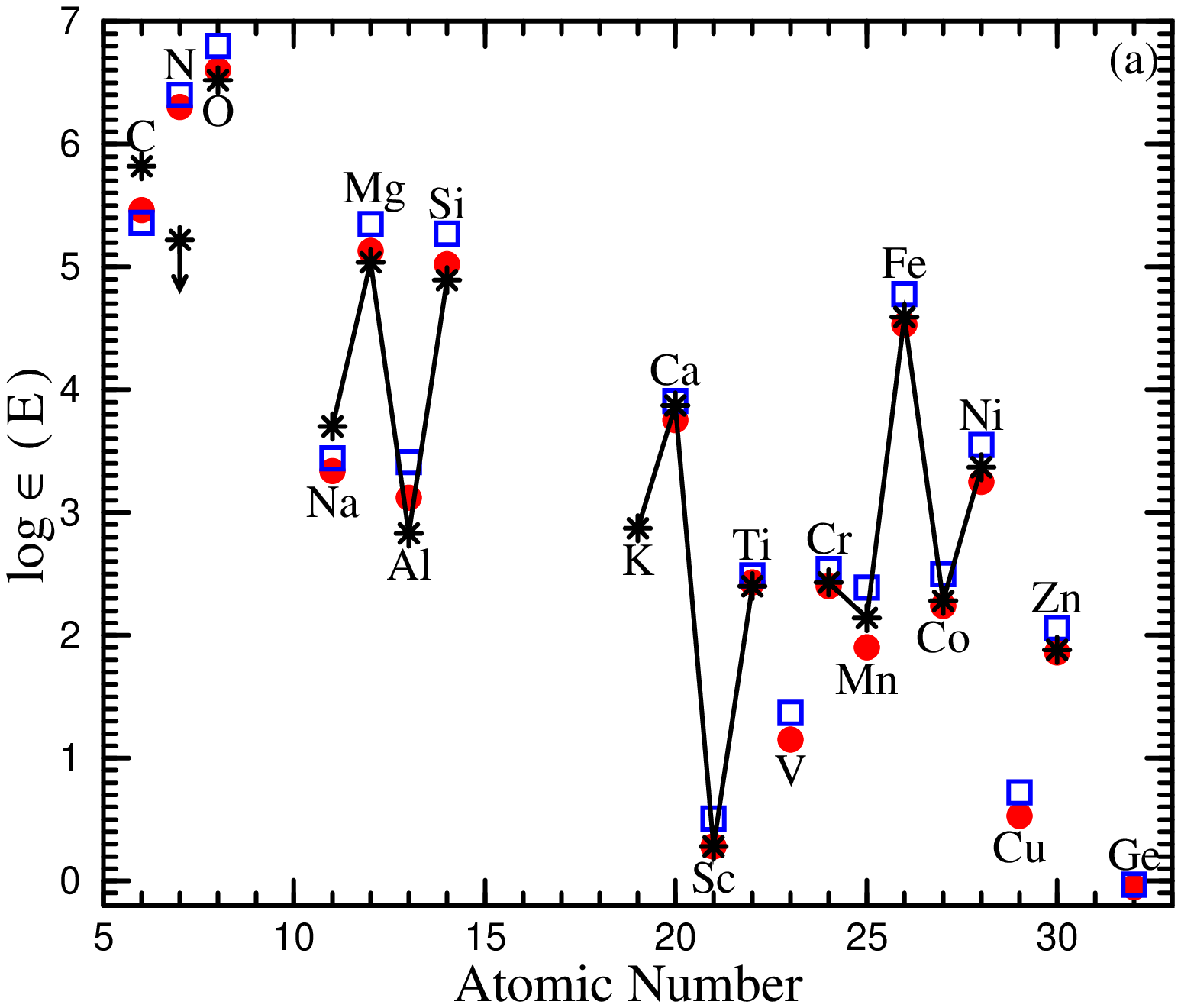}
\includegraphics[scale=0.4]{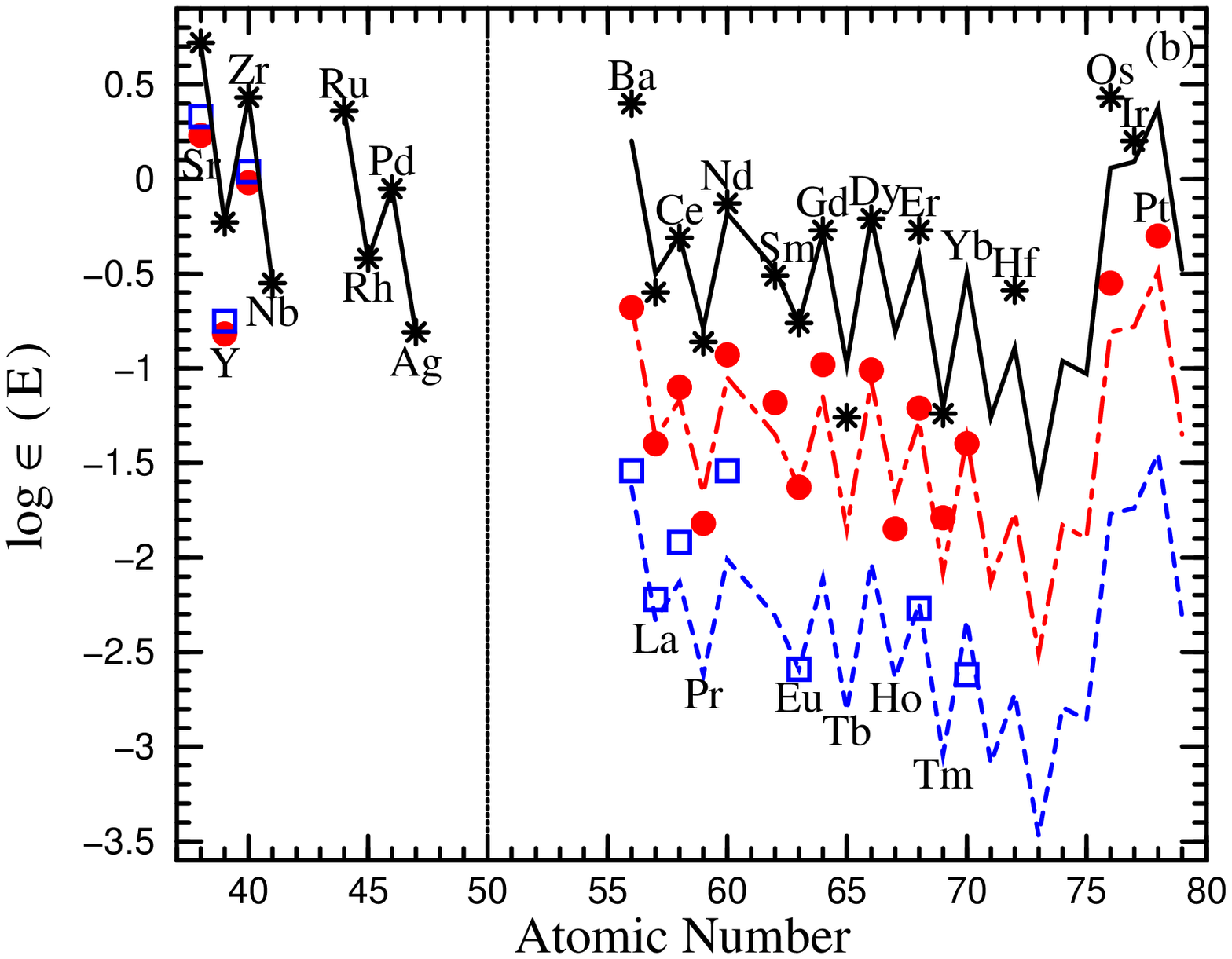}
\caption{Data on the elements from C through Pt in CS~31082--001 (asterisks;
\citealt{hill}), HD~115444 (filled circles), and HD~122563 (squares; \citealt{westin})
with [Fe/H]~$=-2.9$, $-2.99$, and $-2.74$, respectively.
(a) The values of $\log\epsilon({\rm E})\equiv\log({\rm E/H})+12$
for the elements from C through Ge. The data
on CS~31082--001 are connected by solid line segments as a guide.
The downward arrow at the asterisk for N indicates an upper limit. 
Note that the available abundances 
for the elements from O through Ge are almost indistinguishable
for the three stars. (b) The $\log\epsilon$ values for 
the elements from Sr through Pt. The data for
CS~31082--001 in the region to the left of the 
vertical dotted line are again connected by solid line segments as a guide. 
In the region to the right of the vertical dotted line, the data on
the heavy $r$-elements are compared with the solid, dot-dashed, and 
dashed curves, which are the solar $r$-pattern \citep{arlandini} translated to
pass through the Eu data for CS~31082--001, HD~115444, and
HD~122563, respectively. Note the general agreement between the
data and the solid and dot-dashed curves. There is a range of 
$\sim 2$~dex in the abundances of the heavy $r$-elements for the
three stars shown.}
\label{fig-decouple1}
\end{center}
\end{figure}

\begin{figure}[h]
\begin{center}
\includegraphics[scale=0.3, angle=270]{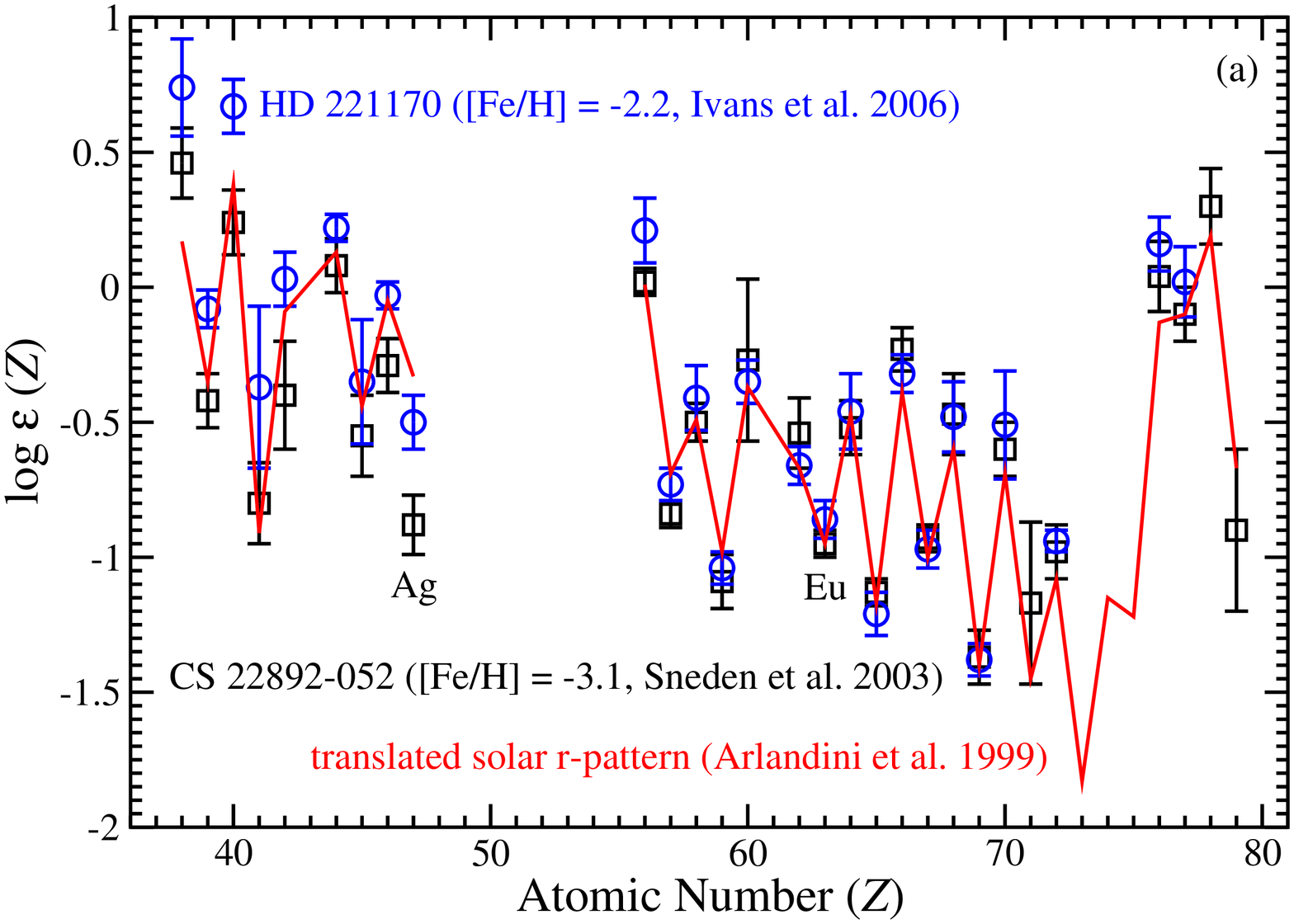}
\includegraphics[scale=0.3, angle=270]{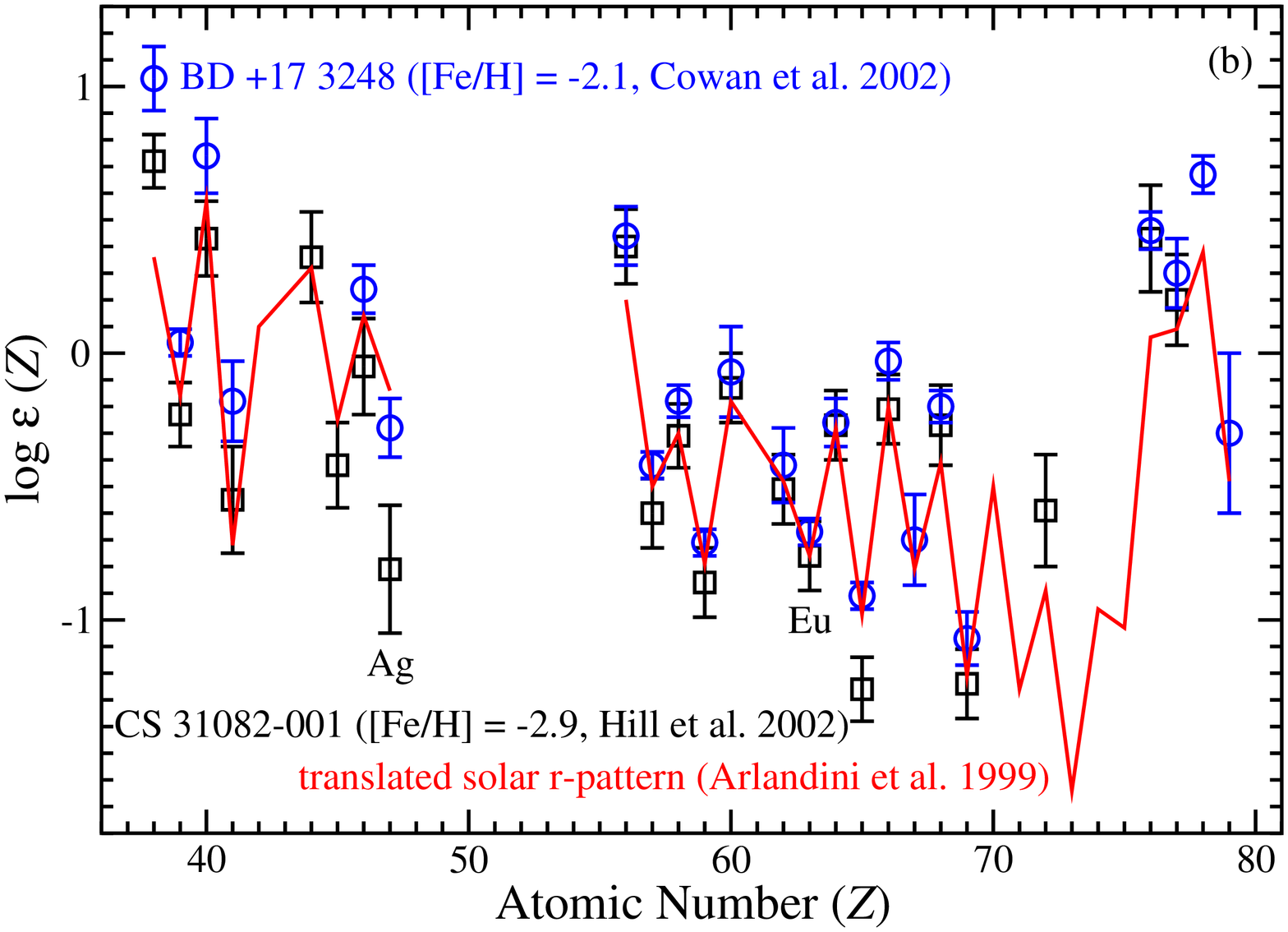}
\caption{Comparison of abundances in two pairs of metal-poor stars.
(a) CS~22892--052 with [Fe/H]~$=-3.1$ (squares with error bars;
\citealt{sneden}) and HD~221170 with 
[Fe/H]~$=-2.2$ (circles with error bars; \citealt{inese}). 
(b) CS~31082--001 with [Fe/H]~$=-2.9$
(squares with error bars; \citealt{hill}) and 
BD~$+17^\circ3248$
with [Fe/H]~$=-2.1$ (circles with error bars; \citealt{cowan}).
The curves in (a) and (b) give the solar ``$r$''-pattern 
\citep{arlandini} translated
to pass through the Eu data for CS~22892--052 and
CS~31082--001, respectively. Note that each pair of stars
have nearly identical abundances of the heavy $r$-elements
closely following the solar $r$-pattern but the (Fe/H) values differ
by a factor of 8 and 6 for the pair in (a) and (b), respectively.
In addition, for an element E in the group from Na through Zn,
the abundance ratio (E/Fe) is the same for the four stars shown 
in (a) and (b) when observational errors are taken into account.}
\label{fig-decouple2}
\end{center}
\end{figure}

Extensive observations of the abundances of a wide range of
elements in metal-poor stars provide further guidance to the search 
of the source for the heavy $r$-nuclei. Figure~\ref{fig-decouple1}b
shows the data on the elements from Sr through Pt for CS~31082--001
(asterisks; \citealt{hill}), HD~115444 (filled circles), and HD~122563 
(squares; \citealt{westin}). The curves
to the right of the vertical dotted line represent the solar $r$-pattern
translated to pass through the Eu data. It can be seen that the data
on the heavy $r$-elements from Ba through Pt in CS~31082--001
and HD~115444 closely follow the solar $r$-pattern. There are fewer 
data on the heavy $r$-elements in HD~122563 and this star will be
discussed in more detail in \S\ref{sec-con}. Note that the levels of 
enrichment in the heavy $r$-elements differ by a factor of $\sim 100$ 
for the three stars shown in Fig.~\ref{fig-decouple1}b. However, their
abundances of the elements from C through Ge are essentially identical
(e.g., [Fe/H]~$\approx -3$) as shown in Fig.~\ref{fig-decouple1}a.
This appears to require that the major source for the heavy $r$-nuclei
produce none of the elements from C through Ge including Fe
\citep{qw02,qw03,qw07}. This decoupling of the production of
the heavy $r$-nuclei from that of the elements from C through Ge
is further demonstrated by comparing CS~22892--052 
([Fe/H]~$=-3.1$; \citealt{sneden})
with HD~221170 ([Fe/H]~$=-2.2$; \citealt{inese})
and CS~31082--001 ([Fe/H]~$=-2.9$) with BD~$+17^\circ 3248$
([Fe/H]~$=-2.1$; \citealt{cowan}). As shown in Fig.~\ref{fig-decouple2},
the stars in either pair have nearly the same abundances of the heavy
$r$-elements, which again closely follow the solar $r$-pattern. 
However, as shown by the data on these stars 
\citep{hill,cowan,sneden,inese} and reflected by their [Fe/H] values,
the abundances of the elements 
between O and Ge differ by a factor of $\sim 8$ and 6 for the former 
and latter pair, respectively. Therefore, the decoupling between the
heavy $r$-nuclei and the elements from C through Ge appears to be
complete: the major source for the former group of nuclei produces
very little of the latter while the major source for the latter produces 
very little of the former. 

The elements from C through Zn are produced between the core and
the H envelope by explosive burning during a core-collapse SN or by
hydrostatic burning during the pre-SN evolution. Stars of $>11\,M_\odot$
develop Fe cores surrounded by extensive shells of Si, O, C, and He.
Consequently, Fe core-collapse SNe from these stars are the major 
source for the elements from C through Zn in the early Galaxy. The
decoupling between these elements and the heavy $r$-nuclei
discussed above then strongly suggests that such SNe are not
the source for the heavy $r$-nuclei. In contrast, stars of 
$\sim 8$--$11\,M_\odot$ develop degenerate O-Ne-Mg cores, 
at least some of which eventually collapse to produce SNe
(e.g., \citealt{nomoto2,nomoto3,iben}).
Models of O-Ne-Mg core-collapse SNe show that
the total amount of material ejected from between the core and the H
envelope is only $\sim 0.01$--$0.04\,M_\odot$
\citep{mayle,kitaura}, much smaller than
the $\sim 1\,M_\odot$ for Fe core-collapse SNe. Thus, O-Ne-Mg
core-collapse SNe contribute very little to the elements from C through Zn.
The decoupling between these elements and the heavy $r$-nuclei
can then be explained by attributing the heavy $r$-nuclei to 
such SNe as argued in \citet{qw02,qw03,qw07}.

O-Ne-Mg core-collapse SNe were also proposed as
the source for the heavy $r$-nuclei based on
considerations of Galactic chemical evolution by
other studies (e.g., \citealt{mathews,ishimaru}).
This proposal is supported by the model for $r$-process
nucleosynthesis presented in \citet{ning}. 
Unlike previous models based on
assumed extremely neutron-rich ejecta (e.g.,
\citealt{wheeler,wanajo}), this new model relies on
the SN shock that rapidly accelerates through
the surface C-O layers of the O-Ne-Mg core due to the steep density
fall-off in these layers. This gives rise to fast
expansion of the shocked ejecta on dynamic timescales of 
$\tau_{\rm dyn}\sim 10^{-4}$~s.
Together with an entropy of $S\sim 100$ and an initial electron fraction 
of $Y_e\sim 0.495$ (e.g., for a composition of ${^{13}{\rm
C}}:{^{12}{\rm C}}:{^{16}{\rm O}}\sim 1:3:3$ by mass), this fast
expansion enables an $r$-process to occur in the shocked ejecta,
producing nuclei with $A>130$ through the actinides.
To further test this model requires two lines of important studies: (1)
calculating the evolution of $\sim 8$--$11\,M_\odot$ stars to
determine the pre-SN conditions of O-Ne-Mg cores, especially the
neutron excess and density structure of the surface layers; and (2)
simulating the collapse of such cores and the subsequent shock
propagation to determine the conditions of the shocked surface
layers. As these layers contain very little mass, simulations with
extremely fine mass resolutions are required to demonstrate the fast
expansion of shocked ejecta that is the key to the
production of the heavy $r$-nuclei in the above model.

\section{Conclusions}
\label{sec-con}
As summarized in Table~\ref{tab}, 
considerations of chemical evolution of heavy elements
in the early Galaxy based on stellar observations and current
understanding of stellar models show that (1) Fe core-collapse SNe 
from progenitors of $>11\,M_\odot$ are the major source for the elements 
from C through Zn in the early Galaxy; (2) O-Ne-Mg core-collapse SNe
from progenitors of $\sim 8$--$11\,M_\odot$ are the major source for the 
heavy $r$-nuclei with $A>130$; and (3) as both Fe and O-Ne-Mg 
core-collapse SNe can produce protoneutron stars that emit neutrinos
to drive winds, both kinds of SNe can produce the elements from Sr through
Ag by CPRs during the $\alpha$-process. The neutrino-driven winds
in some Fe core-collapse SNe (e.g., those producing heavier protoneutron
stars) may even produce $r$-nuclei with $A\sim 130$ (corresponding to
the elements Te, I, and Xe), 
but stellar observations cannot help us to identify the
sources for these nuclei as they are inaccessible to spectroscopic
studies.

\begin{table}[h]
\begin{center}
\caption{Stellar Sources for Heavy Elements}\label{tab}
\begin{tabular}{lcc}
\hline & $>11\,M_\odot^a$ & $\sim 8$--$11\,M_\odot^b$\\
\hline 
C to Zn ($A<70$) & yes & no\\
Sr to Ag ($A\sim 88$--110)& yes & yes\\
$r$-nuclei ($A\sim 130$) & maybe & ?\\
$r$-nuclei ($A>130$) & no & yes\\
\hline
\end{tabular}
\medskip\\
$^a$Progenitors of Fe core-collapse SNe.\\
$^b$Progenitors of O-Ne-Mg core-collapse SNe.\\
\end{center}
\end{table}

\begin{figure}[h]
\begin{center}
\includegraphics[scale=0.3, angle=270]{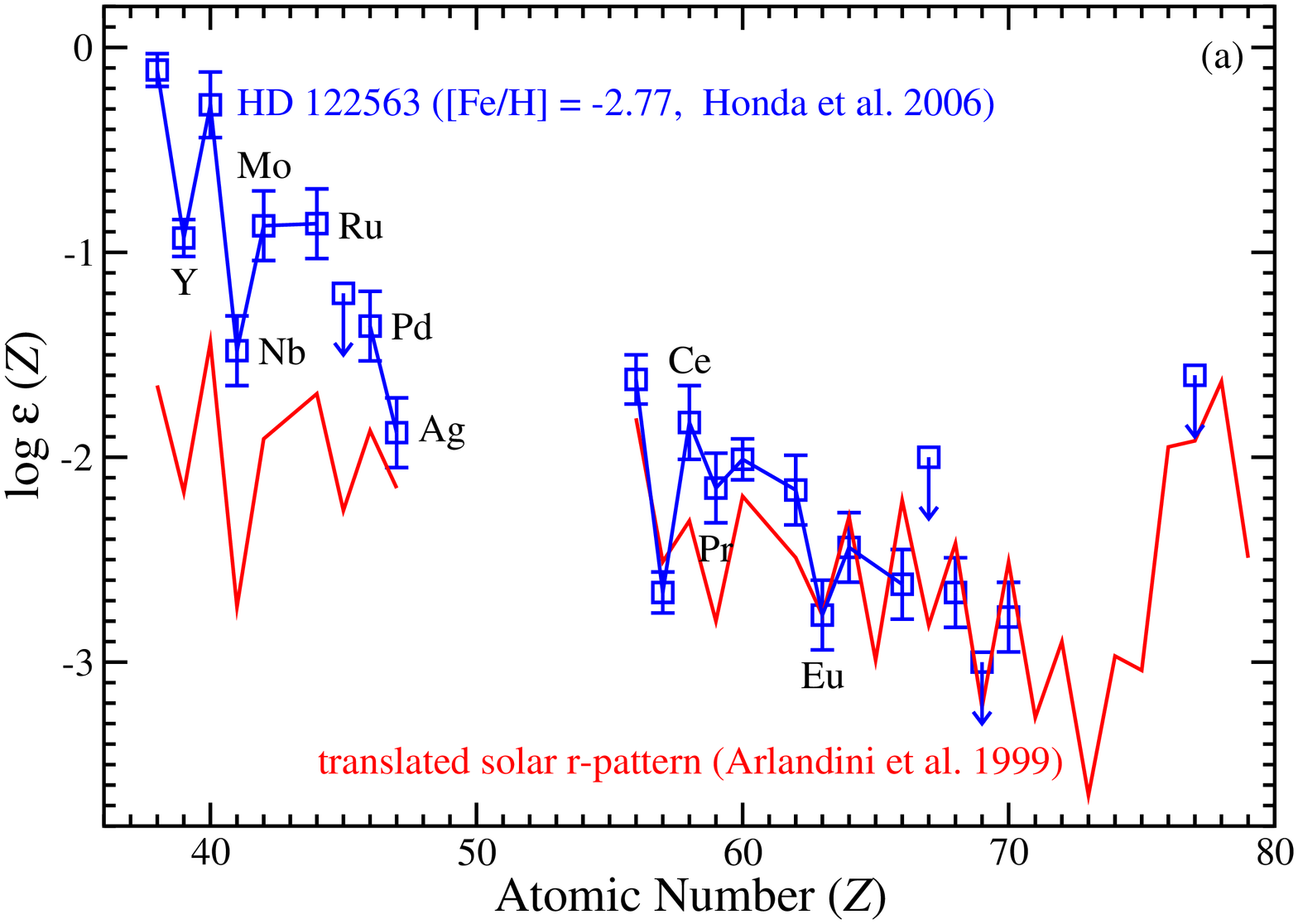}
\includegraphics[scale=0.3, angle=270]{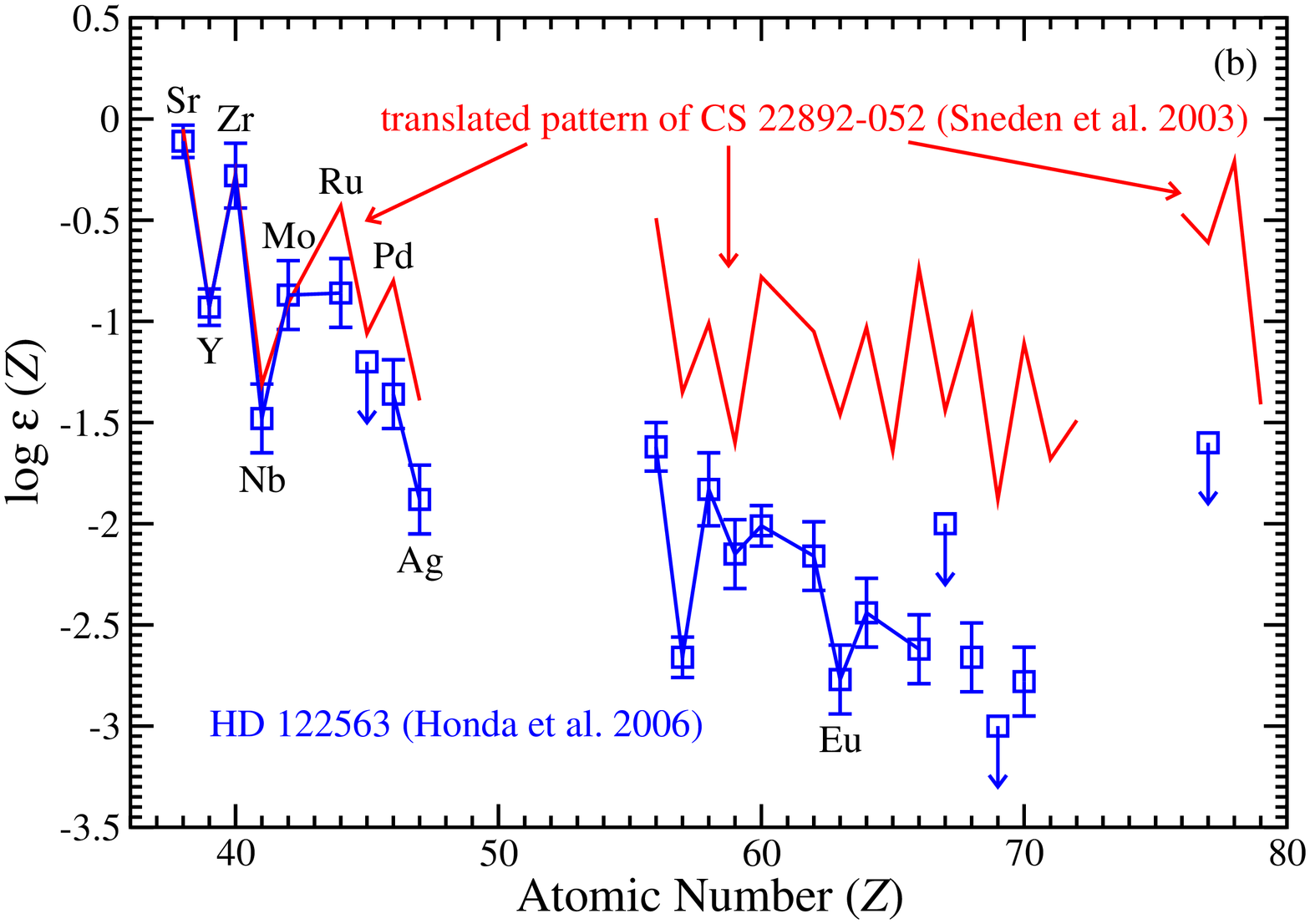}
\caption{(a) Data on HD~122563
(squares with error bars linked by line segments; \citealt{honda}) 
compared with  the solar ``$r$''-pattern
translated to pass through the Eu data (curves labeled as such).
Squares with downward arrows indicate upper limits. 
The abundance pattern of the
heavy $r$-elements [Ba ($Z=56$) and above] in HD~122563
shown in (a) exhibits substantial differences from the solar 
``$r$''-pattern, especially for Ce and Pr. In addition,
HD~122563 has much larger proportions of the elements from 
Sr through Ag relative to the heavy $r$-nuclei as compared to the 
solar ``$r$''-pattern.
(b) Comparison of the data on HD~122563
(squares with error bars linked by line segments; \citealt{honda}) 
with those on CS~22892--052 (curves labeled as such; 
\citealt{sneden}) normalized to the same $\log\epsilon({\rm Y})$ 
as for HD~122563.}
\label{fig-hl}
\end{center}
\end{figure}

Based on the above attribution, one may try to identify
two templates for the overall production of heavy elements
by Fe and O-Ne-Mg core-collapse SNe, respectively. 
The squares with error bars in Fig.~\ref{fig-hl}a
show the data on the elements from Sr through Ir in HD~122563
\citep{honda}. These data are linked
by line segments to guide the eye.
The solar ``$r$''-pattern translated to pass through the Eu data
is shown as the curves labeled so in Fig.~\ref{fig-hl}a.
It can be seen that the data on the heavy $r$-elements Ce, Pr,
Nd, and Sm lie significantly above the translated solar ``$r$''-pattern
while those on the elements from Sr through Ag lie far above it. This
suggests that the source responsible for the abundances in
HD~122563 is not a major contributor to the heavy $r$-elements
but mainly produces the elements from Sr through Ag. Consequently,
the overall abundance pattern in this star including the elements
from C through Zn may be taken as representative of the yields of
Fe core-collapse SNe.  For convenience, this pattern will be 
referred to as the $L$-pattern (dominated by lighter nuclei).
As Fe core-collapse SNe are the major source for Fe in the
early Galaxy, the Fe abundance in a metal-poor star can be used
along with the $L$-pattern to identify the absolute contributions  
from such SNe to all the elements in this star.

As shown in Figs.~\ref{fig-decouple1}
and \ref{fig-decouple2}, a number of metal-poor stars exhibit
a highly regular abundance pattern of the heavy $r$-elements
that is essentially identical to the solar $r$-pattern. 
The data on one such star, CS~22892--052, which are shifted to 
pass through the Y data for HD~122563 for comparison with the 
latter star, are shown as the curves labeled so in 
Fig.~\ref{fig-hl}b. The Fe abundance in CS~22892--052 indicates
that Fe core-collapse SNe contributed very little to its abundances 
of Sr and heavier elements. Figure~\ref{fig-hl}b shows that if the 
source responsible for the abundances of these elements in 
CS~22892--052 produces the elements from Sr through Ag at a
similar level to an Fe core-collapse SN, then this source must
be the major source for the heavy $r$-elements. Consequently,
the abundance pattern of Sr and heavier elements in 
CS~22892--052 can be taken as representative of the yields of
O-Ne-Mg core-collapse SNe. For convenience, this pattern will be 
referred to as the $H$-pattern (dominated by heavier nuclei).
As O-Ne-Mg core-collapse SNe are the major source for the
heavy $r$-element Eu, the Eu abundance in a metal-poor star 
can be used along with the $H$-pattern to identify the absolute 
contributions from such SNe to all the elements in this star.

\begin{figure}[h]
\begin{center}
\includegraphics[scale=0.3, angle=270]{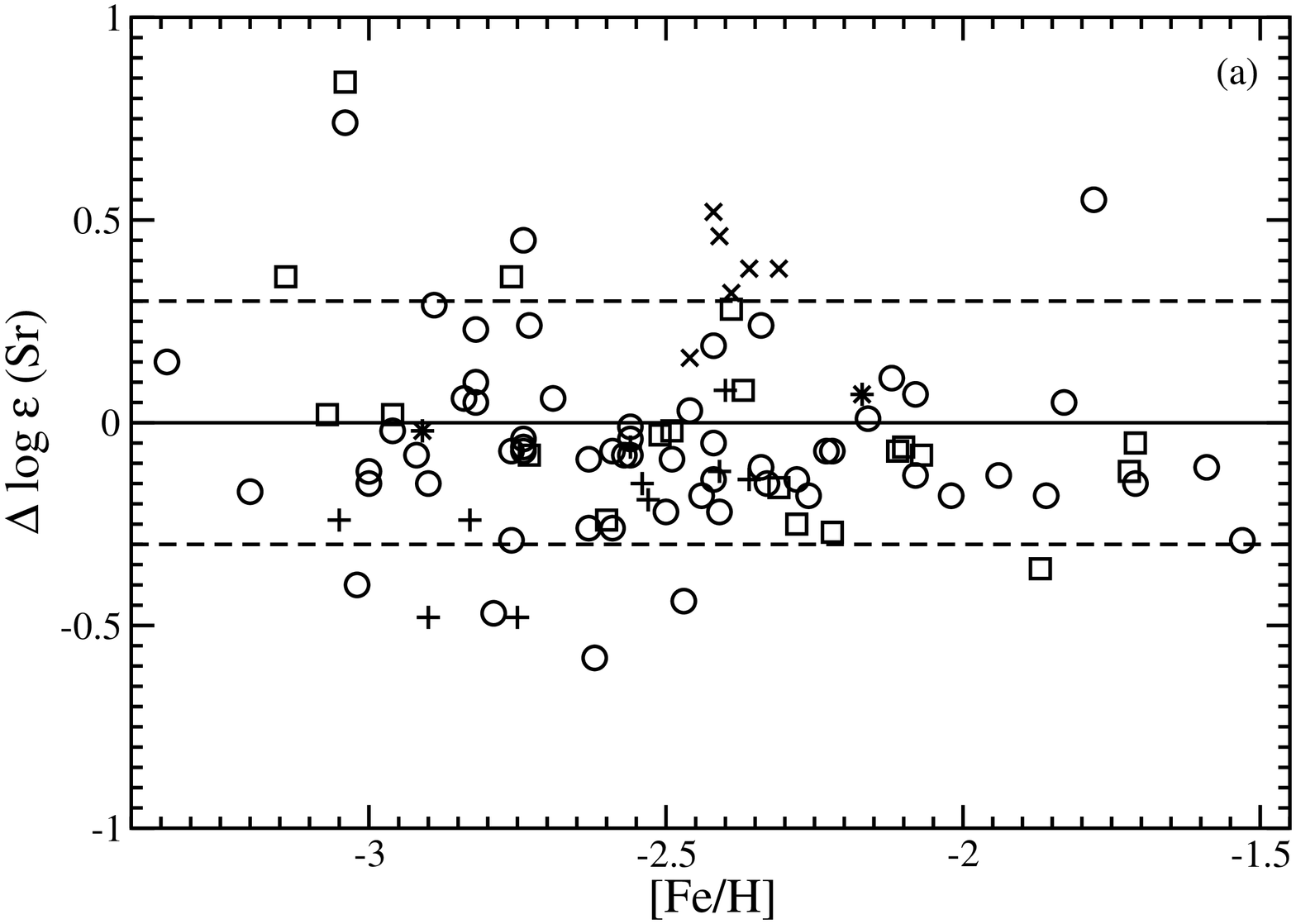}
\includegraphics[scale=0.3, angle=270]{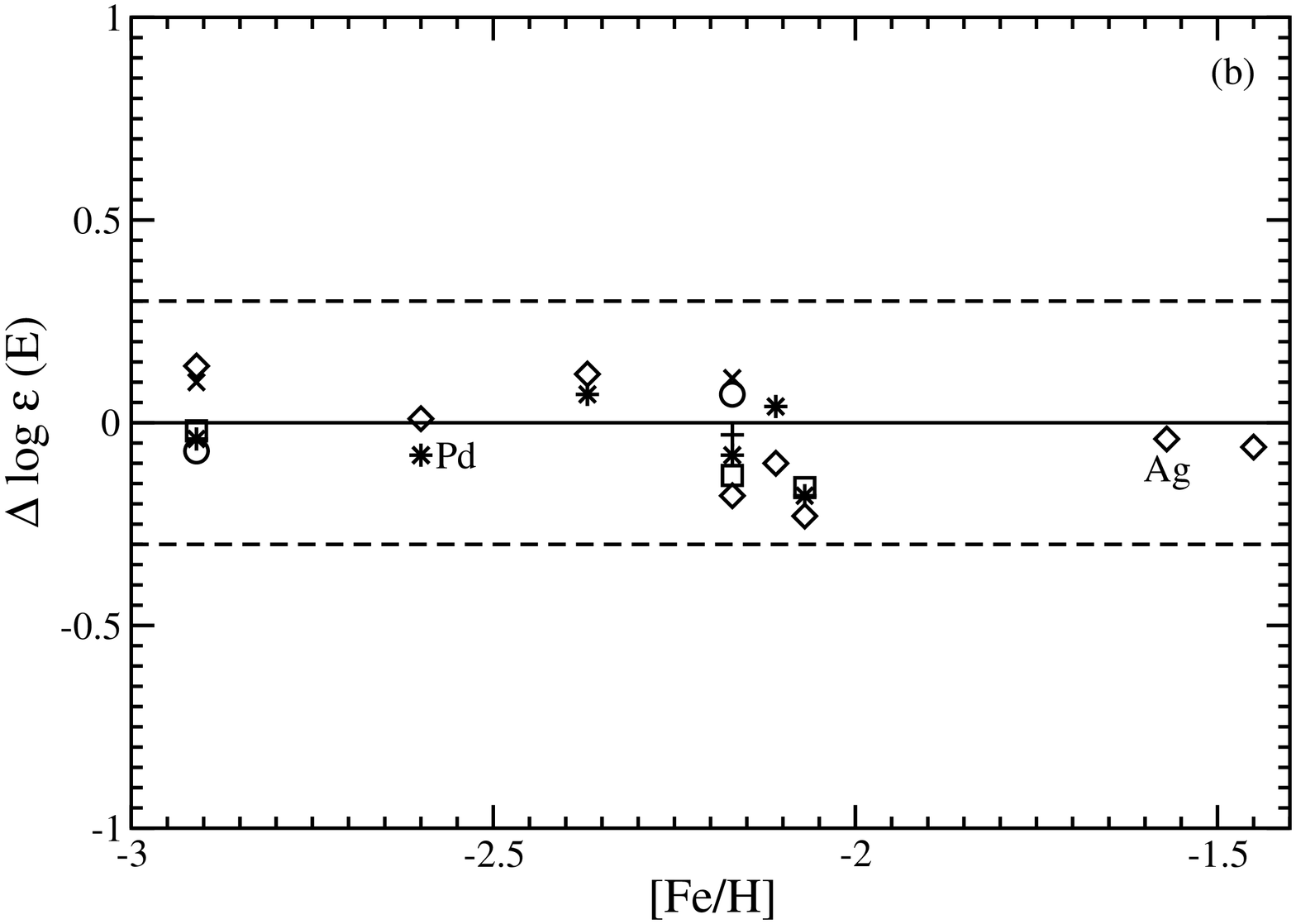}
\caption{Comparison of the two-component model based
on Eq.~(\ref{eq-hl}) with
the data for a large sample of stars. (a) The difference 
between the calculated abundance of Sr
and the observed value is shown in terms of
$\Delta\log\epsilon({\rm Sr})\equiv
\log\epsilon_{\rm cal}({\rm Sr})-\log\epsilon_{\rm obs}({\rm Sr})$
as a function of [Fe/H]. The calculation uses only the observed
Fe and Eu abundances. The symbols represent the data sets 
as follows: squares for \citet{johnson}, pluses for \citet{aoki},
circles for \citet{barklem}, crosses for \citet{otsuki}, and
asterisks for \citet{hill} and \citet{inese}.
(b) Comparison between the model
and data on Nb (squares), Mo (pluses), Ru (circles), Rh (crosses),
Pd (asterisks), and Ag (diamonds). 
The limited data available
are taken from \citet{crawford,hill,cowan,johnson}; and \citet{inese}.
It can be seen from these examples that the elemental 
abundances in a metal-poor star are 
rather well estimated from the model using only the Fe 
and Eu abundances to identify the contributions from two different
sources to the star.}
\label{fig-cpr}
\end{center}
\end{figure}

Based on the above discussion, the abundance of element E
in a metal-poor star can be determined as \citep{qw07}
\begin{equation}
\left(\frac{\rm E}{\rm H}\right)=
\left(\frac{\rm E}{\rm Fe}\right)_L\left(\frac{\rm Fe}{\rm H}\right)+
\left(\frac{\rm E}{\rm Eu}\right)_H\left(\frac{\rm Eu}{\rm H}\right),
\label{eq-hl}
\end{equation}
where (E/Fe)$_L$ and (E/Eu)$_H$ are the yield ratios
representing the $L$-pattern and the $H$-pattern taken from the
data on HD~122563 \citep{honda} and CS~22892--052 
\citep{sneden}, respectively. As an example,
the abundances of Sr for a large 
number of metal-poor stars are calculated from Eq.~(\ref{eq-hl})
using the observed Fe and Eu abundances and compared 
with the data in terms of $\Delta\log\epsilon({\rm Sr})\equiv
\log\epsilon({\rm Sr})_{\rm cal}-\log\epsilon({\rm Sr})_{\rm obs}$
in Fig.~\ref{fig-cpr}a. It can be seen that the calculated Sr
abundances are within 0.3 dex of the data for the majority of
the stars. With the limited data available for the elements
Nb, Mo, Ru, Rh, Pd, and Ag, even better agreement between the 
calculated and observed abundances is obtained for these 
elements as shown 
in Fig.~\ref{fig-cpr}b. In conclusion, the two-component model
based on Eq.~(\ref{eq-hl}) appears to provide a very
good description of the abundances in metal-poor stars. It
remains to be seen whether self-consistent ab initio models of stellar 
evolution and nucleosynthesis can show that the $L$-pattern and the
$H$-pattern are indeed characteristic of the yields of Fe and O-Ne-Mg
core-collapse SNe, respectively.

\section*{Acknowledgments}
This work was supported in part by the US Department of Energy under 
grant DE-FG02-87ER40328.


\begin{thebibliography}{}
\bibitem[Aoki et al.(2005)]{aoki}
Aoki, W., et al. 2005, ApJ, 632, 611

\bibitem[Argast et al.(2004)]{argast}
Argast, D., Samland, M., Thielemann, F.-K., \& Qian, Y.-Z. 2004, A\&A, 416, 997

\bibitem[Arlandini et al.(1999)]{arlandini}
Arlandini, C., et al. 1999, ApJ, 525, 886

\bibitem[Barklem et al.(2005)]{barklem}
Barklem, P. S., et al. 2005, A\&A, 439, 129

\bibitem[Cayrel et al.(2004)]{cayrel}
Cayrel, R., et al. 2004, A\&A, 416, 1117

\bibitem[Chieffi \& Limongi(2004)]{chieffi}
Chieffi, A., \& Limongi, M. 2004, ApJ, 608, 405

\bibitem[Cowan et al.(2002)]{cowan}
Cowan, J. J., et al. 2002, ApJ, 572, 861

\bibitem[Cowan et al.(2005)]{cowan2}
Cowan, J. J., et al. 2005, ApJ, 627, 238

\bibitem[Crawford et al.(1998)]{crawford}
Crawford, J. L., Sneden, C., King, J. R., Boesgaard, A. M., 
\& Deliyannis, C. P. 1998, AJ, 116, 2489

\bibitem[Freiburghaus et al.(1999a)]{freiburghaus1}
Freiburghaus, C., et al. 1999a, ApJ, 516, 381

\bibitem[Freiburghaus et al.(1999b)]{freiburghaus2}
Freiburghaus, C., Rosswog, S., \& Thielemann, F.-K. 1999b, 
ApJ, 525, L121

\bibitem[Fr\"ohlich et al.(2006a)]{frohlich1}
Fr\"ohlich, C., et al. 2006a, ApJ, 637, 415

\bibitem[Fr\"ohlich et al.(2006b)]{frohlich2}
Fr\"ohlich, C., et al. 2006b, Phys. Rev. Lett., 96, 142502

\bibitem[Fuller \& Meyer(1995)]{fuller}
Fuller, G. M., \& Meyer, B. S. 1995, ApJ, 453, 792

\bibitem[Hill et al.(2002)]{hill}
Hill, V., et al. 2002, A\&A, 387, 560

\bibitem[Hoffman et al.(1997)]{hoffman2}
Hoffman, R. D., Woosley, S. E., \& Qian, Y.-Z. 1997, ApJ, 482, 951

\bibitem[Honda et al. (2006)]{honda}
Honda, S., Aoki, W., Ishimaru, Y., Wanajo, S., \& Ryan, S. G. 2006, 
ApJ, 643, 1180

\bibitem[Ishimaru \& Wanajo(1999)]{ishimaru}
Ishimaru, Y., \& Wanajo, S. 1999, ApJ, 511, L33

\bibitem[Ivans et al.(2006)]{inese}
Ivans, I. I., et al. 2006, ApJ, 645, 613

\bibitem[Johnson \& Bolte(2002)]{johnson}
Johnson, J. A., \& Bolte, M. 2002, ApJ, 579, 616

\bibitem[Kitaura et al.(2006)]{kitaura}
Kitaura, F. S., Janka, H.-T., \& Hillebrandt, W. 2006, A\&A, 450, 345

\bibitem[Mathews et al.(1992)]{mathews}
Mathews, G. J., Bazan, G., \& Cowan, J. J. 1992, ApJ, 391, 719

\bibitem[Mayle \& Wilson(1988)]{mayle}
Mayle, R., \& Wilson, J. R. 1988, ApJ, 334, 909

\bibitem[McLaughlin \& Surman(2005)]{mclaughlin}
McLaughlin, G. C., \& Surman, R. 2005, Nucl. Phys. A, 758, 189

\bibitem[Meyer \& Brown(1997)]{meyer2}
Meyer, B. S., \& Brown, J. S. 1997, ApJS, 112, 199 

\bibitem[Meyer et al.(1992)]{meyer1}
Meyer, B. S., Mathews, G. J., Howard, W. M., Woosley, S. E., \& 
Hoffman, R. D. 1992, ApJ, 399, 656

\bibitem[Meynet et al.(2006)]{meynet}
Meynet, G., Ekstr\"om, S., \& Maeder, A. 2006, A\&A, 447, 623

\bibitem[Ning et al.(2007)]{ning}
Ning, H., Qian, Y.-Z., \& Meyer, B. S. 2007, ApJ, 667, L159

\bibitem[Nomoto(1984)]{nomoto2}
Nomoto, K. 1984, ApJ, 277, 791

\bibitem[Nomoto(1987)]{nomoto3}
Nomoto, K. 1987, ApJ, 322, 206

\bibitem[Otsuki et al.(2006)]{otsuki}
Otsuki, K., Honda, S., Aoki, W., Kajino, T., \& Mathews, G. J. 2006,
ApJ, 641, L117

\bibitem[Pruet et al.(2006)]{pruet2}
Pruet, J., Hoffman, R. D., Woosley, S. E., Janka, H. T., \& Buras, R. 2006,
ApJ, 644, 1028

\bibitem[Pruet et al.(2005)]{pruet1}
Pruet, J., Woosley, S. E., Buras, R., Janka, H. T., \& Hoffman, R. D. 2005,
ApJ, 623, 325

\bibitem[Pruet et al.(2003)]{pruet3}
Pruet, J., Woosley, S. E., \& Hoffman, R. D. 2003, ApJ, 586, 1254

\bibitem[Qian(2000)]{qian2}
Qian, Y.-Z. 2000, ApJ, 534, L67

\bibitem[Qian et al.(1993)]{qian1}
Qian, Y.-Z., et al. 1993, Phys. Rev. Lett., 71, 1965

\bibitem[Qian \& Wasserburg(2002)]{qw02}
Qian, Y.-Z., \& Wasserburg, G. J. 2002, ApJ, 567, 515

\bibitem[Qian \& Wasserburg(2003)]{qw03}
Qian, Y.-Z., \& Wasserburg, G. J. 2003, ApJ, 588, 1099

\bibitem[Qian \& Wasserburg(2007)]{qw07}
Qian, Y.-Z., \& Wasserburg, G. J. 2007, Phys. Rep., 442, 237

\bibitem[Qian \& Woosley(1996)]{qwo}
Qian, Y.-Z., \& Woosley, S. E. 1996, ApJ, 471, 331	

\bibitem[Ritossa et al.(1999)]{iben}
Ritossa, C., Garc\'ia-Berro, E., Iben, Jr., I. 1999, ApJ, 515, 381

\bibitem[Sneden et al.(2003)]{sneden}
Sneden, C., et al. 2003, ApJ, 591, 936

\bibitem[Takahashi et al.(1994)]{takahashi}
Takahashi, K., Witti, J., \& Janka,  H.-T. 1994, A\&A, 286, 857

\bibitem[Thompson(2003)]{thompson2}
Thompson, T. A. 2003, ApJ, 585, L33

\bibitem[Thompson et al.(2001)]{thompson1}
Thompson, T. A., Burrows, A., \& Meyer, B. S. 2001, ApJ, 562, 887

\bibitem[Tominaga et al.(2007)]{tominaga}
Tominaga, N., Umeda, H., \& Nomoto, K. 2007, ApJ, 660, 516

\bibitem[Wanajo(2006)]{wanajo2}
Wanajo, S. 2006, ApJ, 647, 1323

\bibitem[Wanajo et al.(2003)]{wanajo}
Wanajo, S., et al. 2003, ApJ, 593, 968

\bibitem[Wanajo \& Ishimaru(2006)]{wanajo3}
Wanajo, S., \& Ishimaru, Y. 2006, Nucl. Phys. A, 777, 676

\bibitem[Westin et al.(2000)]{westin}
Westin, J., Sneden, C., Gustafsson, B., \& Cowan, J. J. 2000, ApJ,
530, 783

\bibitem[Wheeler et al.(1998)]{wheeler}
Wheeler, J. C., Cowan, J. J., \& Hillebrandt, W. 1998, ApJ, 493, L101

\bibitem[Witti et al.(1994)]{witti}
Witti, J., Janka, H.-T., \& Takahashi, K. 1994, A\&A, 286, 841

\bibitem[Woosley \& Baron(1992)]{woba}
Woosley, S. E., \& Baron, E. 1992, ApJ, 391, 228

\bibitem[Woosley \& Hoffman(1992)]{hoffman1}
Woosley, S. E., \& Hoffman, R. D. 1992, ApJ, 395, 202

\bibitem[Woosley \& Weaver(1995)]{ww95}
Woosley, S. E., \& Weaver, T. A. 1995, ApJS, 101, 181

\bibitem[Woosley et al.(1994)]{woosley}
Woosley, S. E., Wilson, J. R., Mathews, G. J., Hoffman, R. D.,
\& Meyer, B. S. 1994, ApJ, 433, 229

\end{thebibliography}
\end{document}